\documentclass[12pt]{article} 
\usepackage{amsmath,amssymb,bm} 
\usepackage{setspace} 
\usepackage{geometry}
\title{Persistent Hydration and Ionic Mobility: When the Ion Is Not the Only Transported Species}

\author{Biman Bagchi\\
Solid State and Structural Chemistry Unit,\\
Indian Institute of Science, Bengaluru 560012, India\\
Email: \texttt{bbagchi@iisc.ac.in}}

\date{}
\begin{document} 
\maketitle

\begin{abstract}
The transport of small ions such as Li$^{+}$, Na$^{+}$, Mg$^{2+}$, and
Ca$^{2+}$ is central to a wide range of problems in solution chemistry,
biology, electrochemistry, and energy storage, including battery
electrolytes.  Yet, despite more than a century of theoretical work on ionic
mobility and electrolyte transport, the microscopic origin of the mobility
of strongly hydrated ions remains only partially understood.  Their transport
cannot in general be described by ionic size and continuum hydrodynamics
alone.  Small, high-charge-density ions can possess strongly structured
hydration shells whose persistence introduces a dynamical scale absent from
conventional continuum and molecular descriptions of ionic friction.

Li$^{+}$ provides a particularly clear example.  Although its bare radius is
smaller than that of Na$^{+}$, its limiting molar ionic conductivity is
lower, while its first hydration shell is dominated by an approximately
four-coordinate Li(H$_2$O)$_4^{+}$ structure.  Related, and in some cases
even stronger, hydration-shell effects occur for multivalent ions such as
Mg$^{2+}$ and Ca$^{2+}$.

We develop a molecular framework in which the ion, its first hydration shell,
and the outer solvent are treated as three dynamically coupled components.
The local hydration structure and water-exchange dynamics are represented
through a potential of mean force and coordination-dependent kinetics, while
the collective response of the outer solvent is described through molecular
memory functions.  The formulation emphasizes that ionic translation and
first-shell exchange need not occur on the same timescale and that hydration
effects cannot, in general, be represented by an independently additive
friction or by a single enlarged hydrodynamic radius.  The resulting
framework extends conventional theories of ionic friction by explicitly
incorporating the dynamics of the first hydration shell and provides a basis
for describing the crossover from strongly hydrated ions to larger ions for
which the traditional molecular description remains more appropriate.
\end{abstract}

\section{Introduction}

The transport of small ions in water is important far beyond the classical
problem of electrolyte conductivity. Li$^{+}$ is central to electrochemical
energy storage and to the design of aqueous and nonaqueous battery
electrolytes, while Mg$^{2+}$ and Ca$^{2+}$ play fundamental roles in
solution chemistry, biological ion transport, and electrochemical systems.
In all of these applications, ionic motion is inseparable from hydration.
For small, high-charge-density ions, the first hydration shell is not merely
an equilibrium structural feature but can remain associated with the ion over
times long compared with elementary solvent relaxation. A theory of their
mobility must therefore address not only the response of water to a moving
charge, but also the motion of the hydrated ionic complex itself.

The anomalous size dependence of the limiting ionic conductivity of the
alkali-metal ions provides one of the simplest manifestations of this
problem. At 298~K, the limiting molar ionic conductivities of Li$^{+}$ and
Na$^{+}$ are approximately 38.7 and 50.1~S\,cm$^{2}$\,mol$^{-1}$,
respectively. Thus the smaller Li$^{+}$ ion conducts substantially less
respectively. Thus the smaller Li$^{+}$ ion conducts substantially less
efficiently than Na$^{+}$, in direct opposition to the elementary Stokes
expectation
\begin{equation}
\zeta_{\rm Stokes}=6\pi\eta R_i ,
\end{equation}
where $\eta$ is the shear viscosity and $R_i$ is the ionic radius.

The foundations of electrolyte transport theory were laid by Debye and
H\"uckel \cite{debye1923} and Onsager
\cite{onsager1926,onsager1927}. Subsequent electrohydrodynamic theories
recognized that ionic friction contains more than the viscous resistance of
the surrounding liquid. A moving ion perturbs the polarization of the
solvent, and the delayed relaxation of this polarization produces dielectric
friction. Important developments were made by Boyd
\cite{boyd1961}, Zwanzig \cite{zwanzig1963,zwanzig1970}, and Hubbard and
Onsager \cite{hubbard_onsager1977}. These theories accounted
semiquantitatively for much of the nonmonotonic size dependence of alkali-ion
mobility, despite representing the ion essentially as a structureless charged
sphere.

A more microscopic description was developed by Wolynes and by Colonomos and
Wolynes \cite{wolynes1978,colonomos_wolynes1979}, and subsequently by Bagchi
and co-workers
\cite{bagchi1991,biswas_roy_bagchi1995,biswas_bagchi1997,
bagchi1998jcp,bagchi_biswas1998}. In these theories the macroscopic dielectric
response is replaced by ion--solvent direct correlation functions,
wave-vector-dependent solvent dynamics, and memory functions. The motion of
the ion and the relaxation of the surrounding polar liquid are therefore
treated as dynamically coupled processes. These approaches achieved
substantial success in describing limiting ionic mobilities and provided a
microscopic interpretation of dielectric friction. The development and
status of this field have been reviewed by Banerjee and Bagchi
\cite{banerjee_bagchi2019}.

The microscopic interpretation of this agreement, however, now requires
reexamination. Neither the continuum theories nor the existing molecular
mode-coupling descriptions explicitly treat a long-lived first hydration
shell that translates together with the ion. Modern quantum-chemical,
QM/MM, and first-principles simulations show that this omission is especially
important for Li$^{+}$. Born--Oppenheimer QM/MM calculations find a strongly
structured first hydration shell with an approximately tetrahedral
arrangement of four water molecules \cite{tongraar_li1998}. Independent
quasi-chemical and electronic-structure analyses likewise identify a
four-coordinate inner hydration environment as the dominant local structure
of aqueous Li$^{+}$ \cite{rempe_li2000}. Further first-principles studies
confirm the highly structured and nonadditive character of Li$^{+}$
hydration \cite{prasetyo_li2018,wilkins_li2015}.

The dynamical consequence of this local structure is crucial. Li$^{+}$ does
not merely possess an enhanced equilibrium ion--water radial distribution
function; it can undergo translational motion while retaining much of its
first hydration environment. Over an intermediate time window, the moving
entity therefore resembles a transient hydrated complex,
Li(H$_2$O)$_4^{+}$, rather than an isolated bare ion. Molecular simulations
by Muralidharan, Pratt, Chaudhari, and Rempe provide further motivation for
distinguishing the dynamics of this hydrated ``solventberg'' from those of
the bare ionic coordinate \cite{muralidharan_pratt2018}.

The situation for Na$^{+}$ is less extreme. Although a distinct inner
hydration region is present, its coordination distribution is broader and
water exchange is more facile
\cite{rempe_na2001,tongraar_nak1998}. Thus the change from Li$^{+}$ to
Na$^{+}$ involves more than a difference in bare ionic size; it also reflects
a qualitative change in the persistence and localization of the first
hydration environment.

The same physical issue appears even more clearly for multivalent cations.
Mg$^{2+}$ possesses a strongly bound, approximately octahedral first
hydration shell containing six water molecules
\cite{lightstone_mg2001,ikeda_mgca2007}. Ca$^{2+}$ also possesses a
well-developed hydration environment, but its coordination and water exchange
are substantially more labile
\cite{tongraar_ca1997,ikeda_mgca2007}. These two ions provide an especially
useful contrast because their translational motion and first-shell exchange
occur on markedly different timescales. For still more strongly polarizing
ions, such as Al$^{3+}$, many-body and polarization effects become even more
pronounced \cite{hofer_al2005}.

These observations suggest a useful dynamical classification. Strongly
hydrated ions such as Li$^{+}$ and Mg$^{2+}$ require an explicit
first-shell degree of freedom. Na$^{+}$ and Ca$^{2+}$ represent an
intermediate regime in which a distinct hydration shell exists but exchanges
more rapidly. For the larger monovalent ions, particularly K$^{+}$,
Rb$^{+}$, and Cs$^{+}$, the first-shell identity is sufficiently labile that
the conventional ion--solvent description is expected to remain much more
appropriate.

From this viewpoint, the substantial quantitative success of earlier
continuum and molecular theories for strongly hydrated ions must be
interpreted with some caution. A theory may reproduce the zero-frequency
diffusion coefficient while still assigning an incomplete microscopic
mechanism to the transport. The agreement obtained for strongly hydrated ions
may therefore conceal an incomplete description of the transported species:
local hydration effects may enter effectively through static correlations or
renormalized friction scales without explicitly describing the translation of
the hydrated complex. This limitation is expected to be much less serious for
the larger alkali ions, for which a persistent transported hydration shell is
absent or far less important.

The theoretical difficulty is therefore not simply to introduce a larger
``hydrated radius.'' Such a replacement combines molecular structure and
hydrodynamic response without describing their dynamical coupling. Nor is it
generally justified to add an independent hydration friction to the viscous
and dielectric contributions. Earlier memory-function and mode-coupling
treatments have shown that different relaxation channels are dynamically
coupled and cannot always be represented as independent additive frictions
\cite{bagchi1991,biswas_roy_bagchi1995,bagchi1998jcp,
banerjee_bagchi2019}.

A second difficulty concerns representation. The established microscopic
theories of ionic friction are naturally formulated in wave-vector space,
where ion--solvent and solvent--solvent direct correlation functions enter
the mode-coupling vertices. The defining physics of the first hydration
shell, in contrast, is intrinsically local and is most naturally described in
real space through ion--oxygen correlations, coordination-number
distributions, potentials of mean force, and water-exchange kinetics.
A useful theory must therefore connect a local real-space description of the
first shell with the collective wave-vector-space dynamics of the outer
solvent.

In the present work we introduce three coupled dynamical density variables,
\begin{equation}
n_i({\bf r},t), \qquad
\rho_f({\bf r},t), \qquad
\rho_s({\bf r},t),
\end{equation}
where $n_i({\bf r},t)$ denotes the ion density,
$\rho_f({\bf r},t)$ the density associated with first-shell water, and
$\rho_s({\bf r},t)$ the density of the remaining solvent. The distinction
between $\rho_f$ and $\rho_s$ is dynamical rather than chemical: individual
water molecules may exchange between the two populations, and $\rho_f$ is
therefore not separately conserved. The ion--first-shell interaction is
described through a real-space potential of mean force and a shell-persistence
or exchange dynamics, whereas the outer solvent retains the
direct-correlation and collective-dynamics description of the earlier
molecular theories.

The principal objective is to understand how the explicit dynamics of the
first hydration shell modifies ionic transport. For Li$^{+}$, this problem is
reflected directly in the relatively modest limiting conductivity despite its
small bare radius. More generally, the observed mobility results from the
competition among bare-size hydrodynamics, microscopic dielectric relaxation,
and the translation and exchange dynamics of the first hydration shell. The
theory developed below is designed to separate these effects without
representing the hydrated ion by a single phenomenological radius or by an
independently additive hydration friction.


\section{Ion--Solvent Coupled Dynamics} 

\subsection{Three dynamical density variables} 

We denote the ion number density by $n_i({\bf r},t)$,
the density of water molecules belonging to the first
hydration shell by $\rho_f({\bf r},t)$, and the density
of the remaining solvent by $\rho_s({\bf r},t)$.
Here ${\bf r}$ denotes position and $t$ denotes time. The distinction between $\rho_f$ and $\rho_s$ is dynamical rather than chemical. A given water molecule may belong to the first shell at one time and to the outer solvent at a later time. Consequently, $\rho_f$ and $\rho_s$ are not independently conserved. Their sum, however, represents the total solvent density and is conserved in the absence of particle creation or loss. The first shell may be defined geometrically through a cutoff radius $r_c$ chosen near the first minimum of the ion--oxygen radial distribution function. If $r_{ij}(t)$ denotes the instantaneous distance between the ion and the oxygen atom of water molecule $j$, a shell indicator may be introduced as \begin{equation} s_j(t)=\Theta\!\left[r_c-r_{ij}(t)\right]. \end{equation} The instantaneous first-shell coordination number is then \begin{equation} N_f(t)=\sum_j s_j(t). \end{equation} For Li$^{+}$, the equilibrium distribution of $N_f$ is strongly concentrated near four, whereas for larger alkali ions the corresponding distribution is broader and the shell is less persistent. The translational position and velocity of the ion are denoted by \begin{equation} {\bf r}_i(t) \qquad\text{and}\qquad {\bf v}_i(t)=\dot{\bf r}_i(t), \end{equation} respectively. When the first shell remains intact over an appreciable time, it is also useful to introduce the center-of-mass position and velocity of the transient hydrated complex, \begin{equation} {\bf r}_X(t), \qquad {\bf v}_X(t)=\dot{\bf r}_X(t), \end{equation} where $X$ denotes the instantaneous ion--first-shell aggregate. The dynamics of ${\bf v}_i(t)$ and ${\bf v}_X(t)$ need not be identical at short and intermediate times. \subsection{Coupled generalized Langevin equations} The most general linearized description involves a set of coupled memory equations. In symbolic form we write \begin{align} \frac{\partial n_i({\bf r},t)}{\partial t} &= -\int_0^t dt' \int d{\bf r}'\, {\cal K}_{ii}({\bf r},{\bf r}';t-t') \frac{\delta {\cal F}}{\delta n_i({\bf r}',t')} \nonumber\\ &\quad -\int_0^t dt' \int d{\bf r}'\, {\cal K}_{if}({\bf r},{\bf r}';t-t') \frac{\delta {\cal F}}{\delta \rho_f({\bf r}',t')} \nonumber\\ &\quad -\int_0^t dt' \int d{\bf r}'\, {\cal K}_{is}({\bf r},{\bf r}';t-t') \frac{\delta {\cal F}}{\delta \rho_s({\bf r}',t')} +\eta_i({\bf r},t), \label{eq:ion_general} \end{align} \begin{align} \frac{\partial \rho_f({\bf r},t)}{\partial t} &= -\int_0^t dt' \int d{\bf r}'\, {\cal K}_{fi} \frac{\delta {\cal F}}{\delta n_i} -\int_0^t dt' \int d{\bf r}'\, {\cal K}_{ff} \frac{\delta {\cal F}}{\delta \rho_f} \nonumber\\ &\quad -\int_0^t dt' \int d{\bf r}'\, {\cal K}_{fs} \frac{\delta {\cal F}}{\delta \rho_s} + {\cal R}_{f\leftrightarrow s}({\bf r},t) +\eta_f({\bf r},t), \label{eq:first_shell_general} \end{align} and \begin{align} \frac{\partial \rho_s({\bf r},t)}{\partial t} &= -\int_0^t dt' \int d{\bf r}'\, {\cal K}_{si} \frac{\delta {\cal F}}{\delta n_i} -\int_0^t dt' \int d{\bf r}'\, {\cal K}_{sf} \frac{\delta {\cal F}}{\delta \rho_f} \nonumber\\ &\quad -\int_0^t dt' \int d{\bf r}'\, {\cal K}_{ss} \frac{\delta {\cal F}}{\delta \rho_s} - {\cal R}_{f\leftrightarrow s}({\bf r},t) +\eta_s({\bf r},t). \label{eq:solvent_general} \end{align} The arguments of the kernels in Eqs.~(\ref{eq:first_shell_general}) and (\ref{eq:solvent_general}) have been suppressed for compactness. The stochastic forces $\eta_i$, $\eta_f$, and $\eta_s$ are related to the corresponding memory kernels through fluctuation--dissipation relations. The exchange term ${\cal R}_{f\leftrightarrow s}$ transfers density between the first shell and the outer solvent. Its opposite signs in Eqs.~(\ref{eq:first_shell_general}) and (\ref{eq:solvent_general}) guarantee that exchange does not change the total number of solvent molecules. At the simplest kinetic level, \begin{equation} {\cal R}_{f\leftrightarrow s} = -k_{\rm off}\rho_f +k_{\rm on}\rho_s, \label{eq:exchange_simple} \end{equation} where $k_{\rm off}$ and $k_{\rm on}$ denote shell escape and capture rates. Equation~(\ref{eq:exchange_simple}), however, should be regarded only as a minimal representation. For Li$^{+}$, where the first shell possesses a well-defined local structure, a more microscopic description should be based on a coordination-dependent potential of mean force. \subsection{Ion--first-shell coupling through a potential of mean force} The ion--first-shell interaction differs qualitatively from the interaction between the ion and the outer solvent. It is local, strongly structured, and is associated with a distinct coordination shell. We therefore do not represent it by an ordinary ion--solvent direct correlation function. A real-space potential of mean force may be defined from the ion--oxygen radial distribution function as \begin{equation} W_{if}(r) = -k_{\rm B}T\ln g_{if}(r). \label{eq:pmf_radial} \end{equation} For shell exchange, an even more useful variable is the coordination number $N_f$. If $P(N_f)$ denotes its equilibrium distribution, the corresponding coordination free energy is \begin{equation} F_f(N_f) = -k_{\rm B}T\ln P(N_f)+{\rm const}. \label{eq:coordination_free_energy} \end{equation} The minima and barriers of $F_f(N_f)$ determine the stability and exchange kinetics of the first hydration shell. In the Li$^{+}$ case, the dominant minimum is expected near $N_f=4$. Transitions to neighboring coordination states provide a natural route for water exchange. A continuous version of this dynamics may be represented by a Smoluchowski equation in a coordination variable $q$, \begin{equation} \frac{\partial P(q,t)}{\partial t} = \frac{\partial}{\partial q} D_q(q)e^{-\beta F_f(q)} \frac{\partial}{\partial q} \left[ e^{\beta F_f(q)}P(q,t) \right]. \label{eq:coordination_smolu} \end{equation} Equation~(\ref{eq:coordination_smolu}) provides a natural connection between the equilibrium shell structure and its time-dependent escape dynamics. In the present work the equilibrium quantities entering $F_f(q)$ are to be taken from existing quantum calculations, ab~initio simulations, or reliable molecular simulations, while the coupling of this shell dynamics to ionic translation and collective solvent relaxation is developed theoretically. \subsection{Mixed real-space and wave-vector-space description} The central structural feature of the present theory is that different pieces of the dynamics are most naturally described in different representations. The first-shell dynamics is local and is therefore formulated in real space or coordination space through $W_{if}(r)$ and $F_f(N_f)$. The collective outer solvent dynamics is naturally represented in wave-vector space. The resulting coupling scheme may be summarized as \begin{equation} n_i \overset{W_{if},\,F_f}{\longleftrightarrow} \rho_f \overset{c_{fs}(k)}{\longleftrightarrow} \rho_s, \end{equation} together with the direct outer-solvent coupling \begin{equation} n_i \overset{c_{is}(k)}{\longleftrightarrow} \rho_s. \end{equation} Thus the ion--first-shell interaction is described through a potential of mean force, whereas the couplings involving the collective outer solvent are represented through direct correlation functions. The theory therefore requires controlled transformations between local real-space shell structure and collective wave-vector-space solvent response. 


\section{Solvent Dynamics} The dynamics of the outer solvent is taken, in the first stage of the theory, from the molecular description of dipolar-solvent relaxation developed in the earlier Bagchi--Chandra framework. The essential idea is that translational density fluctuations and orientational polarization fluctuations of the solvent are coupled and relax on wave-vector-dependent time scales. The ion perturbs these collective solvent modes through ion--solvent direct correlation functions. We denote the outer-solvent density fluctuation by \begin{equation} \delta\rho_s({\bf k},t) = \rho_s({\bf k},t)-\langle\rho_s({\bf k})\rangle. \end{equation} The corresponding dynamic structure factor is \begin{equation} F_{ss}(k,t) = \left\langle \delta\rho_s({\bf k},0) \delta\rho_s(-{\bf k},t) \right\rangle. \label{eq:Fss} \end{equation} For a polar solvent, density fluctuations alone are insufficient. One must also retain the orientational polarization modes. We denote the appropriate longitudinal orientational correlator schematically by \begin{equation} F_{P}(k,t) = \left\langle P_L({\bf k},0) P_L(-{\bf k},t) \right\rangle. \label{eq:FP} \end{equation} The precise tensorial structure will be introduced later when the full Bagchi--Chandra solvent equations are restored. The outer-solvent contribution to the ionic memory kernel then has the same general form as in microscopic mode-coupling theories, \begin{equation} K_{is}(t) \sim \int \frac{d{\bf k}}{(2\pi)^3}\, {\cal V}_{is}({\bf k})\, F_i(k,t)\, F_s(k,t), \label{eq:outer_kernel} \end{equation} where ${\cal V}_{is}({\bf k})$ denotes an ion--solvent vertex constructed from the appropriate direct correlation functions and angular couplings. Equation~(\ref{eq:outer_kernel}) is deliberately schematic at this stage. Its detailed form will depend on the separation of translational and orientational solvent modes. The new element is that the first-shell water alters both the local force on the ion and the boundary condition seen by the outer solvent. Thus the relevant outer-solvent vertex may not be that of a bare ion. At intermediate times the outer liquid may instead respond to a transient hydrated object, which we denote by $X$, \begin{equation} X = [\,i+\text{first hydration shell}\,]. \end{equation} This suggests that an effective shell-dependent vertex, \begin{equation} {\cal V}_{Xs}({\bf k};N_f), \end{equation} may ultimately be required. The first-shell and outer-solvent contributions cannot in general be treated as independent additive frictions. The total force on the ion may be decomposed microscopically as \begin{equation} {\bf F}_i(t) = {\bf F}_{if}(t) + {\bf F}_{is}(t), \end{equation} where ${\bf F}_{if}$ is the force from first-shell molecules and ${\bf F}_{is}$ is the force from the remaining solvent. The corresponding force autocorrelation contains \begin{align} \left\langle {\bf F}_i(0)\cdot{\bf F}_i(t) \right\rangle &= \left\langle {\bf F}_{if}(0)\cdot{\bf F}_{if}(t) \right\rangle + \left\langle {\bf F}_{is}(0)\cdot{\bf F}_{is}(t) \right\rangle \nonumber\\ &\quad + \left\langle {\bf F}_{if}(0)\cdot{\bf F}_{is}(t) \right\rangle + \left\langle {\bf F}_{is}(0)\cdot{\bf F}_{if}(t) \right\rangle. \label{eq:force_decomp} \end{align} Equation~(\ref{eq:force_decomp}) makes clear why the hydration shell cannot be introduced simply as an additional friction coefficient. The first-shell force, the outer-solvent force, and their cross correlations belong to the same microscopic memory function. The present theory therefore retains the established dipolar-solvent dynamics for the outer liquid while introducing a separate slow first-shell variable. For large and weakly hydrated ions the shell exchange time becomes short, the distinction between $\rho_f$ and $\rho_s$ loses dynamical significance, and the conventional two-component ion--solvent theory should be recovered. For Li$^{+}$, and to a lesser extent Na$^{+}$, the shell variable survives on an intermediate time scale and contributes an additional slow component to the ionic memory. The quantitative implementation of this coupling, including the appropriate static direct correlation functions for the outer solvent and the conversion between real-space first-shell structure and wave-vector-space collective dynamics, will be developed in subsequent sections. 

section{Friction Scales and the Structure of the Calculation} Before constructing the full coupled memory-function theory, it is useful to identify the relevant friction scales and estimate their magnitudes. The simplest reference contribution is the hydrodynamic or Stokes friction, \begin{equation} \zeta_{\rm Stokes} = C\,\eta R_i, \label{eq:zeta_stokes} \end{equation} where $\eta$ is the shear viscosity of the solvent and $R_i$ is the ionic radius. We use $R_i$ rather than $r$ for the ion radius because ${\bf r}$ is reserved throughout for the spatial field variable. The symbol $\sigma$ will be used for the characteristic solvent molecular diameter. For a spherical particle with stick boundary conditions, \begin{equation} C=6\pi, \end{equation} whereas ideal slip boundary conditions give \begin{equation} C=4\pi. \end{equation} Although the microscopic boundary condition appropriate to a small molecular ion is not obvious, most earlier treatments of ionic mobility employed the Stokes value $C=6\pi$. We therefore use \begin{equation} \zeta_{\rm Stokes} = 6\pi\eta R_i \label{eq:zeta_stokes_6pi} \end{equation} as our reference hydrodynamic scale, while the $4\pi$ value provides a useful measure of the uncertainty associated with the boundary condition. The corresponding Stokes--Einstein diffusion coefficient is \begin{equation} D_{\rm Stokes} = \frac{k_{\rm B}T}{\zeta_{\rm Stokes}} = \frac{k_{\rm B}T}{6\pi\eta R_i}. \label{eq:D_stokes} \end{equation} Equation~(\ref{eq:D_stokes}) immediately demonstrates the central difficulty of the alkali-ion problem. If the ion were described only by its bare radius, decreasing $R_i$ would monotonically decrease the friction and increase the diffusion coefficient. Li$^{+}$ should therefore be the most mobile ion of the alkali series. Experimentally, however, the limiting ionic conductivity does not continue to increase monotonically as the bare ionic radius decreases. In particular, the conductivity changes only weakly, and in fact decreases, on going from Na$^{+}$ to Li$^{+}$. This departure from the Stokes prediction requires microscopic dynamical contributions. In the earlier molecular theories these arise from the coupling of the moving ion to translational and orientational fluctuations of the surrounding polar solvent. We denote the characteristic microscopic friction associated with this sector by \begin{equation} \zeta_{\rm mic}. \end{equation} In the present theory there is an additional dynamical scale associated with the persistent first hydration shell. We denote the corresponding first-shell memory scale symbolically by \begin{equation} \zeta_f. \end{equation} The notation $\zeta_f$ is introduced only as a convenient measure of the magnitude of the first-shell contribution. It should not be interpreted as an independent friction that can simply be added to the Stokes and microscopic frictions. Indeed, one of the important lessons of earlier kinetic mode-coupling theories is that the hydrodynamic and microscopic transport mechanisms cannot in general be combined through the naive relation \begin{equation} \zeta_{\rm tot} \neq \zeta_{\rm Stokes} + \zeta_{\rm mic} + \zeta_f. \label{eq:no_additive_friction} \end{equation} The separation of microscopic and hydrodynamic dynamics arises from a projection of the full memory kernel onto different classes of collective modes. In the Sj\"ogren--Sj\"olander and related formulations, this leads instead to an inverse-friction or additive-diffusion structure of the schematic form \begin{equation} D = D_{\rm mic} + D_{\rm hyd} = k_{\rm B}T \left[ \frac{1}{\zeta_{\rm mic}} + \frac{1}{\zeta_{\rm hyd}} \right], \label{eq:additive_diffusion} \end{equation} with the precise identification of the two contributions depending on the memory-function decomposition. The new first-shell dynamics must therefore be introduced before the final transport coefficients are combined. In particular, the first hydration shell of Li$^{+}$ should modify the microscopic memory kernel itself rather than appear as a phenomenological friction added at the end of the calculation. The central theoretical task is thus to determine how the persistent first-shell degree of freedom modifies the microscopic ion--solvent memory while remaining consistent with the established microscopic-to-hydrodynamic crossover. For order-of-magnitude estimates it is nevertheless useful to introduce an effective friction obtained directly from a measured or calculated diffusion coefficient, \begin{equation} \zeta_{\rm eff} = \frac{k_{\rm B}T}{D}. \label{eq:zeta_eff} \end{equation} Comparison of $\zeta_{\rm eff}$ with $\zeta_{\rm Stokes}$ provides an immediate measure of the importance of nonhydrodynamic molecular effects. At room temperature the shear viscosity of water is approximately \begin{equation} \eta \simeq 0.89\times10^{-3}\ {\rm Pa\,s}. \end{equation} Using representative bare ionic radii gives Stokes friction coefficients of order $10^{-12}\ {\rm kg\,s^{-1}}$. For Li$^{+}$, with $R_i\simeq0.76$~\AA, Eq.~(\ref{eq:zeta_stokes_6pi}) gives approximately \begin{equation} \zeta_{\rm Stokes}^{\rm Li} \simeq 1.3\times10^{-12}\ {\rm kg\,s^{-1}}, \end{equation} corresponding to \begin{equation} D_{\rm Stokes}^{\rm Li} \simeq 3.2\times10^{-9}\ {\rm m^2\,s^{-1}}. \end{equation} Thus the bare-radius Stokes picture predicts a rather large mobility for Li$^{+}$. The observed limiting conductivity is substantially smaller than would follow from this simple estimate. The calculational problem may therefore be organized around three physical dynamical sectors: \begin{equation} \boxed{ \text{hydrodynamic motion}, \qquad \text{microscopic outer-solvent relaxation}, \qquad \text{first-shell dynamics}. } \label{eq:three_sectors} \end{equation} The first contribution is characterized by $\zeta_{\rm Stokes}$, the second by the microscopic memory generated by the ion--outer-solvent coupling, and the third by the persistence and exchange dynamics of the first hydration shell. The first-shell contribution is expected to be most important for Li$^{+}$, smaller for Na$^{+}$, and progressively less significant for K$^{+}$, Rb$^{+}$, and Cs$^{+}$. Importantly, the earlier molecular theories already provided a reasonably successful description of the experimental mobilities. The new theory is therefore not expected to produce an overwhelmingly large correction to the zero-frequency diffusion coefficient. Rather, the first shell may produce a substantial change in the time-dependent microscopic mechanism while yielding only a moderate correction to the final long-time mobility. This observation places an important constraint on the theory. The first-shell dynamics must account for the fact that Li$^{+}$ can migrate over a significant distance while retaining much of its original hydration shell, but the resulting long-time diffusion coefficient must remain close to the experimentally successful scale obtained in the earlier molecular theories. A useful quantity for characterizing this regime is the distance traveled by the ion during the lifetime of the first shell, \begin{equation} \ell_f = \left(6D_i\tau_f\right)^{1/2}, \label{eq:shell_travel_length} \end{equation} where $\tau_f$ is the characteristic persistence or exchange time of the first hydration shell. Comparison of $\ell_f$ with the first-shell radius and with the solvent molecular diameter $\sigma$ provides a direct measure of whether the ion should be regarded as moving together with a transient hydrated complex. When \begin{equation} \ell_f \ll \sigma, \end{equation} the hydration environment relaxes before the ion has moved appreciably and the conventional two-component ion--solvent description should be adequate. In contrast, when \begin{equation} \ell_f \gtrsim \sigma, \end{equation} the ion translates over molecular distances before losing the identity of its first shell. In this regime the motion of the hydrated complex becomes an independent dynamical quantity. We therefore introduce the velocity autocorrelation function of the bare ion, \begin{equation} C_{v,i}(t) = \left\langle {\bf v}_i(0)\cdot{\bf v}_i(t) \right\rangle, \label{eq:vacf_ion} \end{equation} and the corresponding correlation function of the transient hydrated complex, \begin{equation} C_{v,X}(t) = \left\langle {\bf v}_X(0)\cdot{\bf v}_X(t) \right\rangle, \label{eq:vacf_complex} \end{equation} where $X$ denotes the ion together with its instantaneous first-shell water molecules. One of the principal goals of the present theory is to calculate the diffusion of this transient hydrated complex and to establish its relation to the long-time diffusion of the bare ion. The subsequent calculation will therefore proceed by first determining the outer-solvent microscopic memory, then constructing the first-shell dynamical contribution from the ion--shell potential of mean force and shell persistence dynamics, and finally embedding both within a generalized memory-function description that reduces to the established molecular theory when the first-shell lifetime becomes short.


\section{Effective Ionic Memory: Physical Interpretation
and Limitations}

The central idea of the present approach is that the first
hydration shell of a small ion should not be regarded merely
as a static structural feature. For strongly hydrated ions,
particularly Li$^+$, the first-shell water molecules remain
correlated with the ionic motion over appreciable times.
The moving ion therefore interacts with the outer solvent
both directly and indirectly through its surrounding
hydration shell. This introduces a dynamical distinction
between the bare ionic coordinate and the transient
hydrated complex.

The three-variable formulation developed in Sec.~II
provides a simple way of expressing this distinction.
The ion density, first-shell water density, and outer-solvent
density are coupled through generalized memory functions.
The first-shell and outer-solvent populations are not
independent: water molecules exchange between them, and
their collective fluctuations are correlated. Consequently,
the effective ionic memory cannot generally be obtained
by adding a separate hydration friction to the conventional
dielectric and hydrodynamic contributions.

This point can be expressed formally without specifying
the individual memory kernels. After linearization and
Laplace transformation, the coupled dynamical equations
may be written as
\begin{equation}
\mathbf A(k,z)\widetilde{\mathbf X}(k,z)
=
\mathbf b(k,z),
\end{equation}
where $\mathbf A(k,z)$ is the full dynamical operator,
including the memory kernels, thermodynamic couplings,
and shell-exchange kinetics. Partitioning the variables
into the ionic density and the two solvent densities,
we obtain
\begin{equation}
A_{\mathrm{eff}}(k,z)
=
A_{ii}(k,z)
-
\mathbf A_{iB}(k,z)
\mathbf A_{BB}^{-1}(k,z)
\mathbf A_{Bi}(k,z),
\end{equation}
where $B$ denotes the coupled first-shell and outer-solvent
sector.

This relation follows from the formal elimination of the
two solvent variables. The second term describes their
combined dynamical feedback on the ionic density. It
contains both the individual solvent responses and their
mutual coupling. Thus, even when the ion interacts
directly with the two solvent populations through
different microscopic mechanisms, their effects on ionic
motion need not be independent. The reduced operator
provides a compact representation of this coupling,
although its evaluation still requires microscopic
information about the underlying memory functions.

An important feature of the formulation is its connection
to earlier molecular theories of ionic friction. For ions
with rapidly exchanging hydration shells, the distinction
between first-shell and outer-solvent water becomes less
important on the timescale of ionic relaxation. One
therefore expects an appropriate rapid-exchange limit
to recover an effective two-component ion--solvent
description. For Li$^+$, however, the persistence of
the hydration shell may introduce an additional
intermediate-time relaxation process. This process can
alter the frequency dependence of the ionic response
without necessarily producing a large change in the
zero-frequency diffusion coefficient. Such behavior
would be consistent with the considerable success of
earlier molecular theories in reproducing limiting
ionic conductivities.

The present treatment has several limitations. First,
the coupled memory equations are introduced at a
phenomenological level rather than derived from a
complete microscopic projection-operator calculation.
The formal elimination of solvent variables is exact
within the assumed linear equations, but it does not
determine the memory kernels themselves. Second,
the exchange kinetics of first-shell water are represented
by a simplified description. A quantitative theory would
require coordination-dependent exchange rates and
their coupling to ionic translation. Third, the
equilibrium persistence of a coordination structure
must be distinguished from the residence time of
individual water molecules. A hydration shell may
retain approximately the same coordination number
even while its constituent waters exchange.

A further limitation is that the reduced ionic density
operator is not automatically identical to the
mechanical friction kernel governing the ionic velocity.
Establishing that connection requires the appropriate
conservation laws and dynamical response functions.
Similarly, although the conventional molecular theory
is expected to emerge when the first-shell distinction
becomes dynamically irrelevant, this limiting reduction
has not been explicitly demonstrated here.

The purpose of the present work is therefore not to
provide a complete numerical theory of limiting ionic
conductivity. Rather, it is to identify a physically
motivated extension of the established molecular
description, in which the persistent hydration shell
appears as an explicit dynamical component. The
resulting framework separates the local coordination
and exchange dynamics from the collective relaxation
of the outer solvent, while retaining their mutual
coupling. It provides a basis for future quantitative
calculations and, more immediately, a useful physical
interpretation of the distinctive behavior of small,
strongly hydrated ions.


\section{Extension to Divalent Cations:
Mg$^{2+}$ and Ca$^{2+}$}

The physical picture developed above for Li$^+$ becomes
particularly interesting for divalent cations. Mg$^{2+}$
and Ca$^{2+}$ possess stronger ion--water interactions
than the monovalent alkali ions, but their hydration-shell
structures and exchange dynamics differ substantially.
They therefore provide useful examples for examining
the connection between ionic translation and the
persistence of the first hydration shell.

Mg$^{2+}$ possesses a strongly bound, approximately
octahedral first hydration shell containing six water
molecules. First-principles simulations have established
the stability of this local structure [22,23]. Water
exchange from the first shell is exceptionally slow,
with characteristic residence times of order a
microsecond under ambient conditions. Nevertheless,
Mg$^{2+}$ undergoes translational diffusion through
water. Thus, over times shorter than the shell-exchange
time, its motion may be represented approximately as
that of a transient hydrated complex,
Mg(H$_2$O)$_6^{2+}$.

The separation between translational diffusion and
water exchange can be illustrated using the distance
introduced in Eq.~(41),
\begin{equation}
\ell_f = (6D_i\tau_f)^{1/2}.
\end{equation}
For Mg$^{2+}$, taking
$D_i \simeq 0.7\times10^{-9}\,
\mathrm{m^2\,s^{-1}}$
and $\tau_f \simeq 1\,\mu\mathrm{s}$ gives
$\ell_f \simeq 65\,\mathrm{nm}$.
This estimate is much larger than the dimensions of
the first hydration shell. It illustrates how an ion
can undergo substantial translational displacement
while retaining its coordinated water molecules.
The estimate should be interpreted as a characteristic
diffusive displacement, not as the distance traveled
by a permanently rigid molecular complex.

The situation is different for Ca$^{2+}$. Its larger
ionic radius and lower charge density produce a
less tightly bound hydration environment, with a
broader coordination-number distribution and
substantially faster water exchange [23,24].
Consequently, the distinction between the ion and
its instantaneous hydration complex is expected
to become less pronounced on longer timescales.
Ca$^{2+}$ thus provides an intermediate situation
between the strongly persistent hydration shell of
Mg$^{2+}$ and the more rapidly reorganizing
hydration environments of larger monovalent ions.

These observations suggest that the three-variable
description introduced in Sec.~II has a broader
applicability. For Mg$^{2+}$, the first-shell density
represents a relatively persistent dynamical component,
whereas for Ca$^{2+}$ its exchange with the outer
solvent is more rapid. The effective ionic memory
should therefore depend not only on ionic charge
and radius but also on the coordination structure,
shell-exchange kinetics, and correlations between
the first shell and the surrounding solvent.

A quantitative extension to divalent ions would
require their specific ion--water potentials of
mean force, coordination-dependent exchange rates,
and collective solvent response functions.
These calculations are beyond the scope of the
present work. The important conclusion is that
ionic translation and first-shell water exchange
need not occur on the same timescale, and that
their separation may be especially pronounced
for strongly hydrated multivalent ions.


\section{Conclusions}

The anomalous size dependence of ionic conductivity in
aqueous solutions has long provided an important test
of theories of molecular transport. Continuum
electrohydrodynamics and subsequent microscopic
theories have explained many features of ionic
mobility through the coupling of a moving charge to
viscous, structural, and dielectric relaxation of the
surrounding solvent. The present work examines an
additional physical feature that becomes important
for small, strongly hydrated ions: the dynamical
persistence of the first hydration shell.

Our central proposal is to distinguish three coupled
components: the ion, its first-shell water molecules,
and the remaining solvent. The first hydration shell
is described through local coordination structure,
a potential of mean force, and water-exchange
dynamics, while the outer solvent is treated through
collective molecular relaxation functions. This
mixed real-space and wave-vector-space description
extends the established molecular theory without
introducing an independently additive hydration
friction.

An important consequence is that ionic translation
and first-shell water exchange need not occur on
comparable timescales. Li$^+$ can undergo
translational motion while retaining a substantial
part of its approximately four-coordinate hydration
environment. The distinction becomes even more
striking for Mg$^{2+}$, whose strongly bound
octahedral hydration shell can persist over times
much longer than those associated with molecular
translation. Ca$^{2+}$, with its more labile
hydration environment, provides a useful contrast.
These examples suggest that the persistence of
local hydration structure is an important dynamical
variable in its own right.

The formal elimination of the first-shell and
outer-solvent variables yields an effective ionic
dynamical operator containing their combined
feedback. This result illustrates why first-shell
and outer-solvent contributions cannot generally
be regarded as independent friction mechanisms.
However, the reduced density operator is not
automatically the ionic velocity-friction kernel,
and a quantitative calculation still requires
microscopic memory functions, exchange kinetics,
and an appropriate connection to the ionic
transport coefficient.

The present theory is therefore intended as a
physical and statistical-mechanical framework,
rather than a completed quantitative prediction
of ionic conductivities. Its principal contribution
is to identify the coupled dynamics of ionic
translation, hydration-shell persistence, and
collective solvent relaxation as a natural
extension of existing molecular theories.
Future calculations should determine how these
couplings influence frequency-dependent mobility,
time-dependent correlations, and the limiting
ionic conductivity of strongly hydrated ions.


\end{document}